\documentclass[11pt,twoside,letterpaper]{article} 
\usepackage{times,fancyhdr}
\usepackage[dvips]{graphicx}

  

\makeatletter
\def\cleardoublepage{\clearpage\if@twoside \ifodd\c@page\else%
    \hbox{}%
    \thispagestyle{empty}%
    \newpage%
    \if@twocolumn\hbox{}\newpage\fi\fi\fi}
\makeatother

\def\figurename{Figure}
\makeatletter
\renewcommand{\fnum@figure}[1]{\figurename~\thefigure.}
\makeatother

\def\tablename{Table}
\makeatletter
\renewcommand{\fnum@table}[1]{\tablename~\thetable.}
\makeatother

\begin{document}
\title{
{\begin{flushleft} \vskip 0.45in
{\normalsize\bfseries\textit{Chapter~?}}
\end{flushleft}
\vskip 0.45in \bfseries\scshape Entanglement Relativity in the
Foundations of The Open Quantum Systems Theory }}
\author{\bfseries\itshape M. Arsenijevi\' c$^a$\thanks{E-mail address: fajnman@gmail.com},  J. Jekni\'c-Dugi\' c$^b$, D. Todorovi\' c$^a$, M. Dugi\' c$^a$\\
$^a${\it Department of Physics, Faculty of Science, 34000
Kragujevac, Serbia}\\ $^b${\it Department of Physics, Faculty of
Science and Mathematics, 18000 Ni\v s, Serbia}}
\date{}
\maketitle \thispagestyle{empty} \setcounter{page}{1}
\thispagestyle{fancy} \fancyhead{}
\fancyhead[L]{In: Book Title \\
Editor: Editor Name, pp. {\thepage-\pageref{lastpage-01}}} 
\fancyhead[R]{ISBN 0000000000  \\
\copyright~2007 Nova Science Publishers, Inc.} \fancyfoot{}
\renewcommand{\headrulewidth}{0pt}

\vspace{2in}

\begin{center}
\title{
\bfseries\scshape  Abstract}
\end{center}
Realistic many-particle systems dynamically exchange particles
with their environments. In classical physics, small variations in
the number of constituent particles are commonly considered
practically irrelevant. However, in the quantum mechanical
context, such and similar structural variations are generically
taxed due to the so-called Entanglement Relativity. In this paper
we point out difficulties in deriving master equation for a
subsystem of an alternative partition of the closed quantum
system. We find that the Nakajima-Zwanzig projection method cannot
be straightforwardly used to solve the problem. The emerging tasks
and prospects for the consistent foundations are examined.

\bigskip

\vspace{.08in}

\pagestyle{fancy} \fancyhead{} \fancyhead[EC]{ M. Arsenijevi\' c
et al} \fancyhead[EL,OR]{\thepage} \fancyhead[OC]{Entanglement
Relativity in Open Quantum Systems Theory} \fancyfoot{}
\renewcommand\headrulewidth{0.5pt}

\begin{center}
\title{
\bfseries\scshape  Introduction}
\end{center}

Realistic many-particle systems dynamically exchange particles
with their environments and with other systems. This trivial
observation is still largely intact in the foundations of quantum
theory, including quantum measurement, decoherence and the open
quantum systems theory. While intuitively exchange of particles is
completely clear, it sets a blurred separating line between a
system and its environment. Classical physics straightforwardly
tackles such situations e.g. in motion with variable mass or in
the Grand Canonical Ensemble for systems in thermal equilibrium.
However, systematic quantum-mechanical description of such
processes is lacking.

In quantum measurement or decoherence [1,2,3] or more generally in
open quantum systems [4,5,6] theory,  the border line between
'system' and 'environment' is rarely analyzed in depth. We
construe this fact as a symptom of a subtle and hard problem that
is the  subject of this paper.

The aim of this paper is to diagnose the problem that is still
largely unrecognized but nevertheless of the fundamental
importance in the field of open quantum systems and applications.
We hope that, this first step in noticing and identifying the
problem can help in setting outlines of further progress in the
field.

In Section 2 we carefully define the problem. In Section 3 we
generalize our findings in the context of the so-called
projection-method approach. Section 4 is an illustration of our
considerations and their subtlety. Section 5 is conclusion.

\begin{center}
\title{
\bfseries\scshape  The problem}
\end{center}

Planet Earth is constantly bombarded from the outer space.
Provided that the captured-projectile mass is much less than the
mass of the Earth, the variations in the Earth's orbit around the
Sun are practically negligible. Classical statistical physics
routinely accounts for huge stochastic change in the number of
particles in the many-particle systems in thermal equilibrium. The
Grand Canonical Ensemble of the standard classical statistical
mechanics  describes a composite system of $N$ particles with the
variations of the number of particles from the set
$\{0,1,2,3,...,N\}$. There is also no problem with description of
non-equilibriun mesoscopic systems, such as Brownian particle,
which is dressed by the water molecules that constantly stick to
and come off the particle's surface--the particle's 'hydration
shell' known also for large molecules (e.g. protein molecules and
other biopolymers) in a solution.

Except for the 'chaotic' systems, it seems that classical physics
embraces the following rule: Knowledge of dynamics of an
$N$-particle classical system allows {\it straightforward
deduction} of dynamics for the ($N \pm n$)-particle systems under
the same physical conditions (external fields and interactions),
as long as $n \ll N$. More intuitively, it seems that
individuality of many-particle classical systems is not threatened
by tiny changes in the system's separation from the rest of the
world--e.g. Brownian particle with  $n$ attached water molecules
is typically regarded as practically the same system as  with $n'$
($\neq n$) attached water molecules.

However, in the quantum mechanical context, the things stand
differently. To see this, let us consider the Hamiltonian  of
interest in the quantum Brownian model, in which the environment
$E$ monitors the particle's  center of mass position [4] (and
references therein). For simplicity, we refer to one dimensional
system and the environment consisting of non-interacting linear
harmonic oscillators:

\begin{equation}
\hat H = \hat H_S + \hat H_E + \hat H_{SE}
\end{equation}

\noindent where

\begin{eqnarray}
&\nonumber& \hat H_S = \sum_{i=1}^{N_S} {\hat p_i^2\over 2m_i} +
\sum_{i\neq j=1}^{N_S} V(\vert \hat x_i - \hat x_j\vert)
\\&&\nonumber
\hat H_E = \sum_{\alpha=1}^{N_E} \left({\hat p_{\alpha}^2\over
2m_{\alpha}} + {1\over 2} m_{\alpha}\omega_{\alpha}^2\hat
x_{\alpha}^2\right)
\\&&
\hat H_{SE} = \hat X_{CM} \otimes \sum_{\alpha = 1}^{N_E}
\kappa_{\alpha} \hat x_{\alpha}
\end{eqnarray}

\noindent where the Latin indices refer to the $S$ system and the
Greek indices to the $E$ system, with the numbers $N_S$ and $N_E$
of particles in the system and the environment, respectively, with
the pair interactions $V$, while $\hat X_{CM} = \sum_{i=1}^{N_S}
m_i\hat x_i /M$ and $M = \sum_{i=1}^{N_S} m_i$.

Consider now that the $i_{\circ}$th  particle of the $S$ system
becomes a part of the environment. This is, instead of the $S$
system and the environment $E$ there is the new open system $S' =
S \setminus i_{\circ}$ and the new environment $E' = E \cup
i_{\circ}$ which symbolically reads as a structural (not
necessarily dynamical) transformation:

\begin{equation}
\{S, E\} \to \{S', E'\},
\end{equation}

\noindent while the composite system $C$ as a whole remains intact
by the transformation: $S+E = C = S'+ E'$.

Needless to say, the composite system's Hamiltonian eq.(1) remains
intact by the transformation, while it takes another form:

\begin{equation}
\hat H = \hat H_{S'} + \hat H_{E'} + \hat H_{S'E'}
\end{equation}

\noindent where:

\begin{eqnarray}
&\nonumber& \hat H_{S'} = \sum_{i=1}^{N_S-1} {\hat p_i^2\over
2m_i} + \sum_{i\neq j=1}^{N_S-1} V(\vert \hat x_i - \hat x_j\vert)
\\&&\nonumber
\hat H_{E'} = \hat H_E + {\hat p_{i_{\circ}}^2\over
2m_{i_{\circ}}} + {1\over M} \hat x_{i_{\circ}} \otimes
\sum_{\alpha = 1}^{N_E} \kappa_{\alpha} \hat x_{\alpha}
\\&&
\hat H_{S'E'} = {1\over M} \sum_{i=1}^{N_S - 1} m_i\hat x_i
\otimes \sum_{\alpha = 1}^{N_E} \kappa_{\alpha} \hat x_{\alpha} +
\sum_{j=1}^{N_S-1} V(\vert \hat x_{i_{\circ}} - \hat x_j\vert).
\end{eqnarray}

The following simplifications can make  the two models eq.(2) and
eq.(5) similar to each other: (i) for $N_S \gg 1$, the total mass
$M \approx M ' = \sum_{i=1}^{N_S-1} m_i$, (ii) for large $M$ (i.e.
$M'$), one can neglect the last term in $\hat H_{E'}$, or more
generally to introduce the normal coordinates for the new
environment $E'$ such that it consists of mutually non-interacting
quasi-particles [7,8] and (iii) the pair interactions in $\hat
H_{S'E'}$ to consider as a weak perturbation. Then, according to
the classical intuition, the two open systems $S$ and $S'$ and
their dynamics should appear essentially mutually
indistinguishable.

However, formal similarity of the quantum Hamiltonians does not in
general guarantee that reduced dynamics of the $S$ system
straightforwardly implies the reduced dynamics of the $S'$ system.
This subtle point requires careful examination.

Central to derivation of master equations in the density matrix
theory of open quantum systems  is the tensor product:

\begin{equation}
\hat \rho_S(t) \otimes \hat \rho_E,
\end{equation}

\noindent which is an {\it ansatz} known as {\it Born
approximation} [4,5,6] and follows in a systematic way from the
projection of the composite system's state $\hat \rho(t)$ [9,10]:

\begin{equation}
\mathcal{P} \hat \rho(t) = \hat \rho_S(t) \otimes \hat \rho_E,
\end{equation}

\noindent with $\mathcal{P}^2 = \mathcal{P}$ and $\mathcal{Q} =
\mathcal{I} - \mathcal{P}$ such that $\mathcal{Q}^2 =
\mathcal{Q}$, while $\hat \rho_S(t) = tr_E \hat \rho(t)$ carries
all information about the open system $S$. Linearity of
$\mathcal{P}$ excludes the choice $\hat \rho_E = tr_S\hat
\rho(t)$.

Let us suppose that the reduced dynamics for the $S$ system is
well described by a proper master equation, which assumes validity
of eq.(6), i.e. eq.(7). The main observation of this paper is as
follows:

\smallskip

\noindent ($\mathcal{O}$) {\it As distinct from the classical
counterpart, reduced dynamics of the $S$ system cannot in general
be used to derive or deduce dynamics of the  $S'$ system for the
same time interval.}

\smallskip

This observation asserts that a master equation for the $S$ system
cannot be used to deduce/derive master equation for the $S'$
system for the same time interval, $[0,t]$, with the fixed initial
state, $\hat\rho(t=0)$, of the total $C$ system.

In order to justify the ($\mathcal{O}$), we first emphasize the
formal yet substantial distinction between the classical and
quantum state spaces. In classical physics, the 'phase space' of
$N$ particles is  Cartesian product of the $N$ 'phase spaces' for
individual particles. However, in quantum mechanics, the particles
state spaces are not in Cartesian but in tensor product. For the
composite system $C$ decomposed (structured) as $S+E$, the Hilbert
state space $\mathcal{H}$ is the tensor product of the state
spaces for the subsystems:

\begin{equation}
\mathcal{H} = \mathcal{H}_S \otimes \mathcal{H}_E.
\end{equation}

The structural transformation eq. (3) gives rise to
re-factorization:

\begin{equation}
\mathcal{H} = \mathcal{H}_{S'} \otimes \mathcal{H}_{E'}.
\end{equation}

Invariants of the transformation eq. (3) are: (a) the  Hilbert
state space $\mathcal{H}$, (b) the composite system's Hamiltonian
$\hat H$ and (c) the composite system's state in every instant of
time. However, the transformation eq.(3) induces a change in the
form of the Hamiltonian (e.g. eqs. (2) and (5)) as well as
re-factorization eq.(8) $\to$ eq.(9). The transformation eq.(3)
also leads to a change {\it in the form} of the composite system's
instantaneous state. For a pure state in an instant of time $t$,
this is known as {\it Entanglement Relativity}, e.g. the equality:

\begin{equation}
\vert \phi\rangle_S \otimes \vert \chi\rangle_E = \sum_i c_i \vert
i\rangle_{S'}\otimes\vert i\rangle_{E'}.
\end{equation}

Entanglement Relativity (ER) is a recently established (and
rediscovered) [11-16] corollary of  quantum mechanics that
asserts: Virtually every structural transformation that induces
tensor re-factorization also induces a change in amount of quantum
entanglement in the composite system $C$. If the composite
system's state is tensor-product for one structure ($S+E$) it is
of the entangled form for practically all\footnote{ Exceptions to
ER are also known, but do not alter our main point. E.g. the
transformation eq.(3) does not change a  tensor-product state.
However, the more general structural transformations change even
such states and ER applies [16].} the alternative structures
($S'+E'$) of the composite system. ER regards every (pure) state
in every instant in time $t$.

On the basis of ER, it can be shown that also the more general
non-classical correlations, quantified e.g. by 'quantum discord',
are structure-dependent [17]: A tensor-product mixed state for one
structure ($S+E$) acquires a {\it quantum} correlated (entangled
or discordant) form for virtually arbitrary alternative structure
($S'+E'$) of the composite system, e.g.:

\begin{equation}
\hat \rho_S \otimes \hat \rho_E = \sum_i \lambda_i \hat \rho_{S'i}
\otimes \hat \rho_{E'i}, \quad \sum_i \lambda_i = 1.
\end{equation}

Now the classically unknown conditions eqs. (10) and (11)
plausibly justify the ($\mathcal{O}$) statement as follows; the
more rigorous consideration is given in Section 3. The projection
eq.(7), as well as the generalizations given in Section 3, do not
possess any quantum correlations. Equations (6) and (7) are of
exactly the same form as the l.h.s. of eqs. (10) and (11), which
refer to the $S+E$ structure. The r.h.s. of eqs. (10) and (11),
which refer to the $S'+E'$ structure, carry the quantum
correlations that are not accounted for by any generalization of
eq.(7). Hence derivation of master equation that is {\it based} on
eq.(6), i.e. on eq.(7), for the $S$ system is practically never a
simultaneous derivation of the master equation for the $S'$
system. Consequently, the knowledge of dynamics of the $S$ system
does not suffice to conclude much about dynamics of the $S'$
system.

For the probably most relevant class of Markovian open systems,
eqs. (10) and (11) reveal another layer of consideration. The
tensor-product initial state $\hat \rho_S(t=0) \otimes \hat
\rho_E$ is a necessary condition for Markovian dynamics [5,6];
typically, the environment is supposed thermal, $\hat \rho_E =
\exp(-\beta \hat H_E)/ tr_E \exp(-\beta \hat H_E)$, on the inverse
temperature $\beta = 1/k_B T$. Due to eqs. (10) and (11) for
$t=0$, as long as eq.(6) is valid for the $S+E$ structure, it is
practically never fulfilled regarding the alternative  $S'+E'$
structure. Then Markovian dynamics for the $S$ system is virtually
never applicable  for the alternative $S'$ system. More
specifically: even if eq.(5) can be reduced to eq.(2), and even if
the new environment may also be in thermal-equilibrium state,
$\sum_i \lambda_i \hat \rho_{S'i} = \exp(-\beta' \hat H_{E'})/
tr_E (\exp(-\beta' \hat H_{E'}))$, there is initial correlation
for the $S'$ and $E'$ systems. Consequently, the reduced $S'$
system's dynamics is not Markovian [5,6] and is also possibly
non-completely positive [18].

Hence, typically, derivation of the $S'$ system's dynamics has to
be started from the scratch--by setting eq.(6) for the $S'+E'$
structure in an independent derivation of  master equation for the
$S'$ system. To this end, the knowledge about dynamics of the $S$
system is not useful.

\begin{center}
\title{
\bfseries\scshape  On the use of the Nakajima-Zwanzig projection
method}
\end{center}

One may still wonder if, somehow, dynamics of the $S$ system can
be used for drawing conclusions on the $S'$ system's dynamics.
After all, it's just one tiny-particle difference between the two
structures pertaining to eq. (2) and eq. (5). The correlations
present on the r.h.s. of eqs. (10) and (11) are due only to a
single particle denoted $i_{\circ}$. May it be possible to
approximate the r.h.s. of eqs. (10) and (11) by some
tensor-product states?

In certain special cases (e.g., when the use of the
Born-Oppenheimer approximation is allowed), this may be the case.
Nevertheless there still remains open the question as to whether
'tiny correlations' can be safely discarded from considerations in
general. In this paper we are not interested in such subtle and
deep questions. Rather, we consider usefulness of the
Nakajima-Zwanzig projection method [9,10] in regard of the
($\mathcal{O}$) statement.

Our motivation comes from the fact that the Nakajima-Zwanzig and
the related (projection-based) methods provide systematic
introduction of eq.(6) and set the basis for  the up-to-date the
most general methodological basis of the open systems field
[4,5,6]. If the ($\mathcal{O}$) statement remains valid in the
context of the projection-based methods, then it presents a
serious limitation not only to our classical intuition but also to
operational procedures in describing dynamics of the alternate
open systems. Bearing in mind the classical intuition of Sections
1 and 2, in such a case we face yet another non-trivial task in
the context of the problem of transition from quantum to
classical.

Below, we  assume arbitrary bipartitions of a composite system
$C$, $S+E = C = S'+ E'$, i.e.  arbitrary linear canonical
transformations (LCTs) that induce the tensor-product structures
(i.e. tensor re-factorization) of the composite system's Hilbert
state space.

The key idea behind the Nakajima-Zwanzig projection method [9,10]
is presented by eq.(7). It consists of the introduction of a
linear projection operator, $\mathcal{P}$, which acts on the
operators of the state space of the composite system
'system+environment' ($S+E$). If $\hat \rho$ is the density matrix
of the composite system, the projection $\mathcal{P}\hat\rho$ (the
'relevant part' of the composite density matrix) serves to
represent a simplified effective description through a reduced
state of the composite system. The complementary part (the
'irrelevant part' of the composite density matrix), $\mathcal{Q}
\hat\rho = (I - \mathcal{P})\hat\rho$. For the 'relevant part',
$\mathcal{P}\hat\rho(t)$, one derives closed ('autonomous')
equations of motion in the form of integro-differential equation.
The open system's density matrix $\hat\rho_S(t) = tr_E
\mathcal{P}\hat\rho(t)$ is {\it required} to carry {\it all}
 information about the open system $S$, equivalently
$tr_E \mathcal{Q}\hat\rho = 0$.

The Nakajima-Zwanzig projection method assumes a concrete, in
advance chosen and fixed for all time-instants, system-environment
split (a 'structure'), $S+E$. This split is uniquely defined by
the associated tensor product structure  of the composite system's
Hilbert space, $\mathcal{H} = \mathcal{H}_S \otimes
\mathcal{H}_E$. Division of the composite system into 'system' and
'environment' is practically motivated. In principle, the
projection method can equally describe arbitrary
system-environment split i.e. arbitrary factorization of the
composite system's Hilbert space. By definition, different
factorizations introduce different projectors, denoted
$\mathcal{P}$ for the $S+E$ structure, and $\mathcal{P}'$  for
some alternative $S'+E'$ structure of the composite system, such
that $\hat \rho_S = tr_E \mathcal{P}\hat \rho(t)$ carries all
information about the $S$ system (equivalently $tr_E
\mathcal{Q}\hat \rho(t) = 0$), and $\hat \rho_{S'} = tr_{E'}
\mathcal{P}'\hat\rho(t)$ carries all information about the $S'$
system (equivalently $tr_{E'} \mathcal{Q'}\hat \rho(t) = 0$).

The linear projections  can be defined [1,19]: (i)
$\mathcal{P}\hat\rho(t) = (tr_E \hat\rho(t)) \otimes \hat\rho_E$
[for some $\hat\rho_E \neq tr_S \hat\rho$], which is eq.(7), (ii)
$\mathcal{P} \hat\rho(t) = \sum_n (tr_E \hat P_{Sn} \hat\rho(t))
\otimes \hat\rho_{En}$ [with arbitrary orthogonal supports for the
$\hat\rho_E$s], and (iii) $\mathcal{P} \hat\rho(t) = \sum_i (tr_E
\hat P_{Ei} \hat\rho(t)) \otimes \hat P_{Ei}$ [with arbitrary
orthogonal projectors for the $E$ system]; by $\hat P$, we denote
the projectors on the respective Hilbert state (factor) spaces.
The physical context fixes the choice of the projection--e.g. by
an assumption about the initial state. In this paper we stick to
the projection (i), which is by far of the largest interest in
foundations and applications of the open systems theory. As it can
be easily shown, all the projections (i)-(iii) are free from the
quantum correlations (entanglement or discord).

Now we provide the main results of this section that are borrowed
from [20] with the proofs placed in the appendices.

\noindent
 {\bf Lemma 1.} {\it For the most part of the composite system's
 dynamics, validity of}
\begin{equation}
 tr_E \mathcal{Q}\rho(t) = tr_E (\hat \rho(t) -
\mathcal{P}\hat\rho(t)) =  tr_E (\hat\rho(t) - \hat\rho_S(t)
\otimes \hat\rho_E) = 0, \forall{t}.
\end{equation}

\noindent

\noindent {\it implies non-validity of}

\begin{equation}
 tr_{E'} \mathcal{Q}\hat\rho(t) = tr_{E'} (\hat\rho(t) -
\rho_S(t) \otimes \hat\rho_E) = 0, \forall{t},
\end{equation}

\noindent {\it  and vice versa}.

Lemma 1 reveals that the information 'irrelevant part' of a
projected  state  for one structure contains some relevant
information regarding an alternative structure of the composite
system for the most of  time instants $t$. In formal terms: for
the most part of the composite system's dynamics, the projection
$\mathcal{Q}\hat\rho$ ($\mathcal{Q}'\hat\rho$) brings some
information about the open system $S'$ ($S$)--at variance with the
Nakajima-Zwanzig projection idea. Hence
$\partial\mathcal{P}\hat\rho(t)/\partial t$ allows 'tracing out'
regarding only one structure.  If that structure is $S+E$, then
$tr_{E'} \partial\mathcal{P} \hat\rho(t)/\partial t \neq
\partial\hat\rho_{S'}(t)/\partial t$ [as long as $\hat\rho_{S'}(t)
= tr_{E'}\hat\rho(t)$].  This can be seen also from the following
argument, which is not restricted to the projection-based methods.
Tracing out the $E$ system is dependent on, but not equal to, the
tracing out the $E'$ system, and {\it vice versa}. This dependence
follows from the fact that the $S$ and $E$ degrees of freedom are
intertwined with the $S'$ and $E'$ degrees of freedom.
Intuitively: '$tr_E$' (e.g. integrating over the $E$'s degrees of
freedom) {\it partly} encompasses both the $S'$ and the $E'$
degrees of freedom.

\noindent {\bf Lemma 2.} {\it The two structure-adapted
projectors} $\mathcal{P}$ {\it and} $\mathcal{P}'$  {\it do not
mutually commute}

\begin{equation}
[\mathcal{P},\mathcal{P}']\hat \rho(t)\neq 0
\end{equation}

\noindent {\it for the most of the time instants $t$}.

Very much like noncomutativity of quantum observables, Lemma 2
asserts that the projection-based information contents regarding
different structures of a composite system  are mutually {\it
exclusive} for the most of the time instants $t$. Formally, there
is no state $\hat \rho(t)$ of a composite system for which the
equality $\mathcal{P}\hat \rho(t) = \hat \rho(t) =\mathcal{P'}\hat
\rho(t) $ can be fulfilled for arbitrary instant of time $t$.

Lemma 1 and lemma 2 refer to all projection-based methods and
exclude acquisition of information about an open system $S'$ from
the master equation known for the alternative (albeit possibly
similar) open system $S$ in the same instant (or interval) of
time, and {\it vice versa}. In effect, the ($\mathcal{O}$)
statement is  justified and leads to the conclusion that
derivation of master equations  has to be performed  for every set
of the degrees of freedom (i.e. for every open system, $S$, $S'$
etc.) separately, in accord with equations (10) and (11).

\pagebreak

\begin{center}
\title{
\bfseries\scshape  Analysis of the quantum Brownian motion}
\end{center}

In order to illustrate subtlety of our considerations, we stick to
eq.(2) as the Caldeira-Leggett model of quantum Brownian motion
(QBM)  [4,21].

In the Schr\" odinger picture the QBM master equation for the
initial separable state ($\hat \rho(t=0) = \hat \rho_S(t=0)
\otimes \hat \rho_E$) with the environment on temperature $T$:

\begin{equation}
{d\hat\rho_S(t)\over dt} = -{\imath \over\hbar} [\hat H_S,\hat
\rho_S(t)] - {\imath \gamma\over \hbar}[\hat x_S, \{\hat p_S, \hat
\rho_S(t)\}] - {2m\gamma k_B T\over\hbar^2} [\hat x_S, [\hat x_S,
\hat\rho_S(t)]].
\end{equation}

\noindent The curly brackets denote the 'anticommutator', $m$ is
the mass while $\hat x_S$ and $\hat p_S$ are the position and
momentum of the particle  and $\gamma$ is the
 semi-empirical friction coefficient.

Eq.(15) is not of the Lindblad form and hence by definition [5,6]
is not Markovian. Interestingly enough, eq.(15) applies even for
initially correlated state and for arbitrary strength of
interaction in the composite system as well as for arbitrary
'spectral density' (which defines the friction coefficient
$\gamma$)\footnote{See e.g. eq.(4.226) in Ref. [4].}.
Non-Markovianity of eq.(15) is behind its 'robustness', which is
the ultimate basis of the observation of QBM effect for an
alternative structure of the total system $C$ [7].

Now we emphasize the variations offered by eq.(15). First, it is
known that eq.(15) can be transformed in a Lindblad form for
sufficiently high temperature $T$  [4]. Second, for the massive
particle, the second term proportional to $\gamma$ can be
neglected.  This constitutes the 'recoilless' variant of the
Caldeira-Leggett model and  provides the Lindblad-form master
equation:

\begin{equation}
{d\hat\rho_S(t)\over dt} = -{\imath \over\hbar} [\hat H_S,\hat
\rho_S(t)] - {2m\gamma k_B T\over\hbar^2} [\hat x_S, [\hat x_S,
\hat\rho_S(t)]],
\end{equation}

\noindent which is similar with the scattering-decoherence master
equation--see eq.(3.66) in [1].

It is remarkable that already at the level of {\it fixed
structure}, $C=S+E$, we can see non-trivial variations in the form
of master equation and consequently regarding the $S$  system's
dynamics.

Now we refer to certain {\it structural variations} of the
dynamics described by eq.(16). We are aiming at the cases in which
the classical intuition can be justified; see e.g. comments below
eq.(5). In all other cases we do not expect the classical
intuition to be very useful.

Concretely, we are interested in eq.(3) as well as in the opposite
case, i.e. when an environmental particle, denoted
$\alpha_{\circ}$, is joined the $S$  system thus providing a new
open system, $S'' = S \cup \alpha_{\circ}$, and new environment,
$E'' = E\setminus \alpha_{\circ}$; needless to say,
$C=S+E=S'+E'=S''+E''$.\footnote{For examples of the more general
structural transformations see e.g. Refs. [7,16,22,23].} This
situation also describes the 'Schr\" odinger's cat'--the cat
represented by the $S$ system while the $\alpha_{\circ}$th
environmental particle flows out of the radioactive source.

Regarding the structural transformation

\begin{equation}
(S,E) \to (S'', E''),
\end{equation}

\noindent that is accompanied by tensor re-factorization,
$\mathcal{H}_{S}\otimes\mathcal{H}_E \to
\mathcal{H}_{S''}\otimes\mathcal{H}_{E''}$ , the Hamiltonian
eq.(2) i.e. eq.(5) takes the form:

\begin{equation}
\hat H = \hat H_{S''}  + \hat H_{E''} + \hat H_{S''E''}.
\end{equation}

In eq.(18):

\begin{eqnarray}
&\nonumber& \hat H_{S''} = \hat H_S + {\hat
p^2_{\alpha_{\circ}}\over 2m_{\alpha_{\circ}}} + \hat
X_{CM}\otimes \kappa_{\alpha_{\circ}} \hat x_{\alpha_{\circ}}
\\&&\nonumber
\hat H_{E'} = \sum_{\alpha=1}^{N_E-1} \left({\hat
p_{\alpha}^2\over 2m_{\alpha}} + {1\over 2}
m_{\alpha}\omega_{\alpha}^2 \hat x_{\alpha}^2\right)
\\&&
\hat H_{S''E''} = \hat X_{CM} \otimes \sum_{\alpha = 1}^{N_E-1}
\kappa_{\alpha} \hat x_{\alpha}.
\end{eqnarray}

As distinct from eq.(5), eq.(19) is already of the form of eq.(2):
there only appears a new interaction in the system's (in the $S''$
system's) self Hamiltonian\footnote{Rigorously, the new structure
$(S+\alpha_{\circ})+E''$, where the $\alpha_{\circ}$ particle is
not in interaction with $E''$. If we assume that the mass $M''$ of
$S''$ is approximately $M$, the two models eq.(2) and eq.(19)
become practically indistinguishable.}. Therefore, the two
structure variations, eq.(3) and eq.(17), are not mutually
equivalent.

Our starting model is eq.(2), for which we assume the initial
tensor product state and thermal environment $E$:

\begin{equation}
\hat \rho(t=0) = \hat \rho_S \otimes \hat \rho_E = \hat \rho_S
\otimes_{\alpha} {\exp(-\beta\hat H_{\alpha})\over Z_{\alpha}},
\end{equation}

\noindent with the one-particle 'statistical sum' $Z_{\alpha} =
tr_{\alpha} \exp(-\beta\hat H_{\alpha})$.

The interactions generated by the $V$ pair-interactions in the $S$
system suggest correlations in the initial $S$ system's state,
$\hat \rho_S(t=0) = \sum \mu_m \hat \rho^{S'}_m \otimes \hat
\rho^{i_{\circ}}_m$, which gives correlated initial state for the
$S'+E'$ structure:

\begin{equation}
\hat \rho(t=0) = \sum_m \mu_m \hat \rho^{S'}_m \otimes \hat
\rho^{E'}_m \equiv \sum_m \mu_m \hat \rho^{S'}_m \otimes \left(
\hat \rho^{i_{\circ}}_m \otimes \hat \rho^E\right), \quad \sum_m
\mu_m = 1.
\end{equation}

However, at variance with eq.(11)\footnote{This nicely exhibits
the subtlety of ' quantum correlations relativity', eq.(10) and
eq.(11):  for some special states (here: tensor-product states)
and for a special pair of structures (here: $S+E$ and $S''+E''$)
one should not worry about the quantum correlations relativity.
However, the worry remains for almost all other kinds of
re-structuring even for the initial tensor-product state regarding
the starting $S+E$ structure of the total system.}, eq.(20)
directly provides tensor-product initial state  for the $S''+E''$
structure:

\begin{equation}
\hat \rho(t=0) =\hat \rho_{S''} \otimes \hat \rho_{E''} \equiv
\left( \hat \rho_S \otimes {\exp(-\beta\hat
H_{\alpha_{\circ}})\over Z_{\alpha_{\circ}}} \right)
\otimes_{\alpha\neq\alpha_{\circ}} {\exp(-\beta\hat
H_{\alpha})\over Z_{\alpha}}.
\end{equation}

Further we focus on the $S''+E''$ structure.

The projection is defined:

\begin{equation}
\mathcal{P}\hat\rho(t)  = \hat \rho_S(t) \otimes_{\alpha}
{\exp(-\beta\hat H_{\alpha})\over Z_{\alpha}},
\end{equation}

\noindent providing that $\otimes_{\alpha} \exp(-\beta\hat
H_{\alpha})/ Z_{\alpha} \neq tr_S \hat \rho(t)$ (and analogously
for the other structures).

For the $S''+E''$ structure, the projection reads:

\begin{equation}
\mathcal{P''}\hat\rho(t) = \left(tr_{E\setminus
\alpha_{\circ}}\hat\rho(t)\right)\otimes\hat\rho_{E''}.
\end{equation}

In an  instant of time $t>0$, we expect correlations in the $S''$
system, e.g. $\hat \rho_{S''}(t) := tr_{E\setminus
\alpha_{\circ}}\hat\rho(t) = \sum_i \lambda_i(t) \hat
\rho^{S}_{i}(t) \otimes \hat \rho^{\alpha_{\circ}}_i(t)$. Hence
with the use of eq.(24):

\begin{equation}
tr_{E''}{\partial \mathcal{P''}\hat\rho(t)\over \partial t} =
{\partial \over \partial t}\sum_i \lambda_i(t) \hat
\rho^{S}_{i}(t) \otimes \hat \rho^{\alpha_{\circ}}_i(t).
\end{equation}

On the other hand, from eq.(23):

\begin{equation}
tr_{E''} \mathcal{P}\hat\rho(t) = \hat \rho_S(t) \otimes
{\exp(-\beta\hat H_{\alpha_{\circ}})\over Z_{\alpha_{\circ}}},
\end{equation}

\noindent which instead of  eq.(25) gives:

\begin{equation}
tr_{E''}{\partial \mathcal{P}\hat\rho(t)\over \partial t} =
 {\partial \over \partial t} \hat \rho_S(t)\otimes {\exp(-\beta\hat H_{\alpha_{\circ}})\over Z_{\alpha_{\circ}}}.
\end{equation}

The absence of the exact correlations appearing in eq.(25) clearly
illustrates Lemma 1, i.e. implies that
$tr_{E''}\mathcal{Q}\hat\rho(t) \neq 0$, even in this
case\footnote{See eq.(22).} in which eq.(11) is not applicable.

Hence despite the classical similarity, the open systems $S$, $S'$
and $S''$ are subjected to different dynamics. While the $S$
system, as well possibly as the $S''$ system, undergoes a
frictionless Markovian dynamics eq.(16), the $S'$  system may be
expected to be subjected to neither Markovian [5,6] nor
frictionless [4,21] and possibly non-completely-positive [18]
dynamics. Exact master equations for the $S'$ and $S''$ systems
can  follow from the independent e.g. projection-based analysis.
Regarding the $S'$ system, derivation of the master equation
should start from the projection $\mathcal{P}' \hat\rho(t) = \hat
\rho_{S'}(t)\otimes \hat \rho_{E'}$ with the initial state of the
form of eq.(21), while for the $S''$ system, the derivation should
regard eq.(25) with the initial state of the form of eq.(22)--that
does not bring any substantial new observation that is of interest
for the present paper.

\begin{center}
\title{
\bfseries\scshape  Conclusion}
\end{center}

Recently, it was shown [7] that non-similar subsystems of a
composite system can have similar dynamics. In contrast to the
classical intuition of Sections 1 and 2, in this paper we show
that classically indistinguishable many-particle systems can
undergo non-equivalent dynamics. Moreover, the shortcuts in
describing dynamics of the alternative open systems may not even
exist. It is our conjecture that the classical intuition fits with
some special structures of many-particle systems [24] that can
still require certain assumptions, such as e.g. the
Born-Oppenheimer adiabatic approximation. Those results set a new
layer in the long standing problem of the transition from quantum
to classical [1] that will be discussed elsewhere.

Perhaps not surprisingly, our findings are implied by the
classically unknown Quantum Correlations Relativity [11-17]--a
not-yet-fully-appreciated rule of the universally valid quantum
theory--and some other, classically surprising, findings may be
expected.

Currently it appears that description  of the alternative
subsystems dynamics should be performed for every alternate open
system separately. The classical intuition, that similar systems
should bear similar dynamics, appears unreliable.

\bigskip

\begin{center}
\title{
\bfseries\scshape  Acknowledgement}
\end{center}

The work on this paper is financially supported by Ministry of
education, science and technological development, Serbia, grant no
171028 and in part for MD by the ICTP-SEENET-MTP grant PRJ-09
"Strings and Cosmology" in frame of the SEENET-MTP Network.

\begin{center}
\title{
\bfseries\scshape  References}
\end{center}

[1] Joos, E.; Zeh, H. D.; Kiefer, C.; Giulini, D.; Kupsch, J.;
Stamatescu, I.-O. \textit{Decoherence and the Appearance of a
Classical World in Quantum Theory (2nd edition)}; Springer:
Berlin, 2003.

[2] Zurek, W.H. \textit{Rev. Mod. Phys.} 2003, {\it 75}, 715-775.

[3] Schlosshauer, M. \textit{Rev. Mod. Phys.} 2005, {\it 76},
1267-1305.

[4] Breuer, H.-P.; Petruccione, F. \textit{The Theory of Open
Quantum Systems}; Clarendon Press: Oxford, 2002.

[5] Rivas, \' A.; Huelga, S. F. \textit{Open Quantum Systems. An
Introduction}; Springer: Berlin, 2011.

[6] Rivas, \' A.; Huelga, S. F.; Plenio M. B. \textit{Rep. Prog.
Phys.} 2014, {\it 77}, 094001.

[7] Dugi\' c, M.; Jekni\' c-Dugi\' c, J. \textit{Pramana J. Phys.}
2012, {\it 79}, 199-209.

[8] Lim, J.; Tame, M.; Yee, K. H.; Lee, J.-S.; Lee, J. \textit{
New J. Phys.} 2014, {\it 16}, 018001.

[9] Nakajima S. \textit{Prog. Theor. Phys.} 1958, {\it 20},
948-959.

[10] Zwanzig R. \textit{J. Chem. Phys.} 1960, {\it 33}, 1338-1341.

[11] Dur, W.; Vidal, G. \textit{Phys. Rev. A} 2000, {\it 62},
062314.

[12]  Eakins, J.; Jaroszkiewicz G. \textit{J. Phys. A: Math. Gen.}
2003, {\it 36}, 517-526.

[13] Zanardi, P.; Lidar, D. A.; Lloyd, S. \textit{Phys. Rev.
Lett.} 2004, {\it 92}, 060402.

[14] Ciancio, E.; Giorda, P.; Zanardi P.\textit{ Phys. Lett. A}
2006, {\it 354}, 274-280.

[15] Dugi\' c, M.; Jekni\' c, J. \textit{Int. J. Theor. Phys.}
2006, {\it 45}, 2215-2225.

[16] Jekni\' c-Dugi\' c, J.; Arsenijevi\' c, M.; Dugi\' c, M.
\textit{Quantum Structures: A View of the Quantum World}; LAP
Lambert Acad Publ: Saarbr\" ucken, 2013.

[17] Dugi\' c, M.; Arsenijevi\' c, M.; Jekni\' c-Dugi\' c, J.
\textit{Sci. China Phys. Mech. Astron.} 2013, {\it 56}, 732-736.

[18] Brodutch, A.; Datta, D.; Modi, K.; Rivas, \' A.;
Rodriguez-Rosario, C. A. \textit{Phys. Rev. A} 2013, {\it 87},
042301.

[19] Gemmer, J.; Michel, M.; Mahler, G. \textit{Quantum
Thermodynamics, Lecture Notes in Physics};  Springer-Verlag:
Berlin, 2004; Vol. 657.

[20] Arsenijevi\' c, M.; Jekni\' c-Dugi\' c, J.; Dugi\' c, M.
(2013). A Limitation of the Nakajima-Zwanzig projection method.
http://arxiv.org/abs/1301.1005.

[21] Caldeira, A. O.; Leggett, A. \textit{ Physica A} 1983, {\it
121}, 587.

[22] Stokes, A.; Kurcz, A.; Spiller, T. P.;  Beige, A.
\textit{Phys. Rev. A} 2012,    {\it 85}, 053805.

[23] Arsenijevi\' c, M.; Jekni\' c-Dugi\' c, J.; Dugi\' c, M.
\textit{Chin Phys B}
 2013, {\it 22}, 020302.

[24] Arsenijevi\' c, M.; Jekni\' c-Dugi\' c, J.; Dugi\' c, M.
(2013). Zero Discord for Markovian Bipartite Systems.
http://arxiv.org/abs/1204.2789.

\begin{center}
\title{
\bfseries\scshape  Appendix A: Proof of Lemma 1}
\end{center}

Given eq.(12), i.e. $tr_E \mathcal{Q}\hat\rho(t) = 0, \forall{t}$,
we investigate the conditions that should be fulfilled in order
for eq.(13), i.e. $tr_{E'} \mathcal{Q}\hat\rho(t) = 0,
\forall{t}$, to be fulfilled. The $\mathcal{Q}$ projector refers
to the $S+E$, not to the $S'+E'$ structure. Therefore, in order to
calculate $tr_{E'}\mathcal{Q}\hat\rho(t)$, we use ER. We refer to
the projection (i), Section 3, in an instant of time:

\begin{equation}
\label{eq.10}
 \mathcal{P} \hat\rho = (tr_E \hat\rho) \otimes \hat\rho_E.
\end{equation}

\noindent A) Pure state $\hat\rho = \vert \Psi \rangle\langle \Psi
\vert$, while, due to eq.(12), $tr_E \mathcal{Q} \vert \Psi
\rangle\langle \Psi \vert = 0$.

We consider the pure state presented in its (not necessarily
unique) Schmidt form

\begin{equation}
\label{eq.11}
 \vert \Psi \rangle = \sum_i c_i \vert i\rangle_S
\vert i\rangle_E,
\end{equation}

\noindent where $\hat\rho_S = tr_E \vert \Psi\rangle\langle
\Psi\vert = \sum_i p_i \vert i\rangle_S\langle i\vert$, $p_i =
\vert c_i\vert^2$ and for arbitrary $\hat\rho_E \neq tr_S \vert
\Psi\rangle\langle\Psi\rangle$. Given $\hat\rho_E = \sum_{\alpha}
\pi_{\alpha} \vert \alpha \rangle_E\langle \alpha \vert $, we
decompose $\vert \Psi\rangle$ as:

\begin{equation}
\label{eq.12} \vert \Psi\rangle = \sum_{i, \alpha} c_i C_{i\alpha}
\vert i\rangle_S \vert \alpha\rangle_E,
\end{equation}

\noindent with the constraints:

\begin{equation}
\label{eq.13}
 \sum_i \vert c_i\vert^2 = 1 = \sum_{\alpha}
\pi_{\alpha}, \sum_{\alpha} \vert C_{i\alpha} \vert^2 = 1,
\forall{i},
\end{equation}

Then

\begin{equation}
\label{eq.14}
 \mathcal{Q} \vert \Psi\rangle\langle \Psi \vert =
\vert \Psi \rangle\langle \Psi \vert - \sum_{i, \alpha} p_i
\pi_{\alpha} \vert i\rangle_S\langle i\vert \otimes \vert \alpha
\rangle_E\langle \alpha \vert.
\end{equation}

We use ER:

\begin{equation}
\label{eq.15}
 \vert i\rangle_S \vert \alpha\rangle_E = \sum_{m,n}
D^{i\alpha}_{mn} \vert m\rangle_{S'} \vert n \rangle_{E'}
\end{equation}

\noindent with the constraints:

\begin{equation}
\label{eq.16}
 \sum_{m,n}  D^{i\alpha}_{mn} D^{i'\alpha'\ast}_{mn}
= \delta_{ii'} \delta_{\alpha\alpha'}.
\end{equation}

With the use of eqs.(\ref{eq.12}) and (\ref{eq.15}),
eq.(\ref{eq.14}) reads:

\begin{equation}
\label{eq.17}
 \sum_{m,m'n,n'}[\sum_{i,i', \alpha, \alpha'} c_i
C_{i\alpha} c_{i'}^{\ast}C_{i'\alpha'}^{\ast} D^{i\alpha}_{mn}
D^{i'\alpha' \ast}_{m'n'} - \sum_{i,\alpha} p_i \pi_{\alpha}
D^{i\alpha}_{mn} D^{i\alpha\ \ast}_{m'n'}] \vert
m\rangle_{S'}\langle m' \vert \otimes \vert n\rangle_{E'}\langle
n'\vert.
\end{equation}

After tracing out, $tr_{E'}$:

\begin{equation}
\label{eq.18}
 \sum_{m,m'} \left.\{  \sum_{i,\alpha,n}
\sum_{i',\alpha'} c_i C_{i\alpha}
c_{i'}^{\ast}C_{i'\alpha'}^{\ast} D^{i\alpha}_{mn}
D^{i'\alpha'\ast}_{m'n}\right. -  \left. p_i \pi_{\alpha}
D^{i\alpha}_{mn} D^{i\alpha\ast}_{m'n} \right.\} \vert
m\rangle_{S'}\langle m'\vert
\end{equation}

Hence

\begin{equation}
\label{eq.19}
 tr_{E'} \mathcal{Q} \vert \Psi\rangle\langle
\Psi\vert = 0 \Leftrightarrow \sum_{i,\alpha,n} [\sum_{i',\alpha'}
c_i C_{i\alpha} c_{i'}^{\ast} C_{i'\alpha'}^{\ast}
D^{i\alpha}_{mn} D^{i'\alpha'\ast}_{m'n} -  p_i \pi_{\alpha}
D^{i\alpha}_{mn} D^{i\alpha\ast}_{m'n}] = 0, \forall{m,m'}.
\end{equation}

Introducing notation, $\Lambda^m_n \equiv \sum_{i,\alpha} c_i
C_{i\alpha} D^{i\alpha}_{mn}$, one obtains:

\begin{equation}
\label{eq.20}
 tr_{E'} \mathcal{Q} \vert \Psi\rangle\langle
\Psi\vert = 0 \Leftrightarrow \\  A_{mm'} \equiv \sum_{n}
[\Lambda^m_n \Lambda^{m'\ast}_n - \sum_{i,\alpha}p_i \pi_{\alpha}
D^{i\alpha}_{mn} D^{i\alpha\ast}_{m'n}] = 0, \forall{m,m'}.
\end{equation}

Notice:

\begin{equation}
\label{eq.21} \sum_m A_{mm} = 0.
\end{equation}

\noindent which is equivalent to $tr \mathcal{Q} \vert \Psi\rangle
\langle \Psi \vert = 0$, see eq.(\ref{eq.14}).

\noindent B) Mixed (e.g. non-entangled) state.

\begin{equation}
\label{eq.22}
 \hat\rho = \sum_i \lambda_i \hat\rho_{Si}\hat\rho_{Ei}, \quad
\hat\rho_{Si} = \sum_m p_{im} \vert \chi_{im}\rangle_S
\langle\chi_{im} \vert,   \\ \hat\rho_{Ei} = \sum_n \pi_{in}\vert
\phi_{in}\rangle_E\langle \phi_{in}\vert,
\end{equation}

In eq.(\ref{eq.22}), having in mind eq.(\ref{eq.10}), $tr_E
\mathcal{Q} \hat\rho = 0$, while $tr_E\hat\rho = \sum_p \kappa_p
\vert \varphi_{p}\rangle_S\langle \varphi_{p}\vert$, and
$\hat\rho_E = \sum_q \omega_q \vert \psi_{q}\rangle_E\langle
\psi_{q}\vert \neq tr_S \hat\rho$.

Constraints:

\begin{equation}
\label{eq.23}
 \sum_i \lambda_i = 1 = \sum_p \kappa_p = \sum_q
\omega_q, \quad \sum_{m} p_{im} = 1 = \sum_n \pi_{in}, \forall{i}.
\end{equation}

Now we make use of ER and, for comparison, we use the same basis
$\{\vert a\rangle_{S'} \vert b \rangle_{E'}\}$

\begin{equation}
\label{eq.24}
 \vert \chi_{im}\rangle_S \vert \phi_{in} \rangle_E =
\sum_{a,b} C^{imn}_{ab} \vert a \rangle_{S'} \vert b \rangle_{E'},
 \\ \vert\varphi_p\rangle_S \vert \psi_q\rangle_E = \sum_{a,b}
D^{pq}_{ab} \vert a \rangle_{S'} \vert b \rangle_{E}.
\end{equation}

Constraints:

\begin{equation}
\label{eq.25}
 \sum_{a,b} C^{imn}_{ab} C^{im'n'\ast}_{ab} =
\delta_{mm'} \delta_{nn'}, \quad \sum_{a,b} D^{pq}_{ab}
D^{p'q'\ast}_{ab} = \delta_{pp'} \delta_{qq'}.
\end{equation}

So

\begin{eqnarray}
&\nonumber& \label{eq.26} \mathcal{Q}\hat\rho = \hat\rho -
(tr_E\hat\rho) \otimes \hat\rho_E = \sum_{a,a',b,b'}
\{\sum_{i,m,n} \lambda_i p_{im} \pi_{in} C^{imn}_{ab}
C^{imn\ast}_{a'b'} -
\\&&
\sum_{p,q} \kappa_p \omega_q D^{pq}_{ab} D^{pq\ast}_{a'b'}\} \vert
a \rangle_{S'}\langle a'\vert \otimes \vert b\rangle_{E'} \langle
b'\vert.
\end{eqnarray}

Hence

\begin{equation}
\label{eq.27}
 tr_{E'} \mathcal{Q} \hat\rho = 0 \Leftrightarrow
\Lambda_{aa'} \equiv \sum_{i,m,n,b} \lambda_i p_{im} \pi_{in}
C^{imn}_{ab} C^{imn\ast}_{a'b}-  \sum_{p,q,b} \kappa_p \omega_q
D^{pq}_{ab} D^{pq\ast}_{a'b} = 0, \forall{a,a'}.
\end{equation}

Again, for $a=a'$:

\begin{equation}
\label{eq.28}
 \sum_a \Lambda_{aa} = 0,
\end{equation}

\noindent as being equivalent with $tr \mathcal{Q} \hat\rho = 0$,
see eq.(\ref{eq.26}).

Validity of eq.(13) assumes validity of eq.(\ref{eq.20}) for pure
and of eq.(\ref{eq.27}) for mixed states. Both eq.(\ref{eq.20})
and eq.(\ref{eq.27}) represent the sets of simultaneously
satisfied equations. We do not claim non-existence of the
particular solutions to eq.(\ref{eq.20}) and/or to
eq.(\ref{eq.27}), e.g. for the finite-dimensional systems.
Nevertheless, we want to emphasize that the number of states they
might refer to is apparently negligible compared to the number of
states for which this is not the case. For instance, already for
the fixed $a$ and $a'$, a small change e.g. in $\kappa$s (while
bearing eq.(\ref{eq.23}) in mind) undermines equality in
eq.(\ref{eq.27}).

Quantum dynamics is continuous in time. Provided eq.(12) is
fulfilled, validity of eq.(13) might refer {\it only} to a special
set of the time instants. So we conclude: for the most part of the
open $S'$-system's dynamics, eq.(13) is not fulfilled. By
exchanging the roles of eq.(12) and eq.(13) in our analysis, we
obtain the reverse conclusion, which completes the proof. Q.E.D.

\begin{center}
\title{
\bfseries\scshape  Appendix B: Proof of Lemma 2}
\end{center}

The commutation condition, $[\mathcal{P}, \mathcal{P}']\hat\rho(t)
= 0, \forall{t}$. With the notation $\hat\rho_P(t) \equiv
\mathcal{P}\hat\rho(t)$ and $\hat\rho_{P'}(t) \equiv
\mathcal{P}'\hat\rho(t)$, the commutativity reads:
$\mathcal{P}\hat\rho_{P'}(t) = \mathcal{P}'\hat\rho_P(t),
\forall{t}$. Then $\mathcal{P}\hat\rho_{P'}(t) =
tr_E\hat\rho_{P'}(t) \otimes \hat\rho_E = \hat\rho_S(t) \otimes
\hat\rho_E$, while $\mathcal{P}'\hat\rho_{P}(t) =
tr_{E'}\hat\rho_P(t) \otimes  = \sigma_{S'}(t) \otimes
\sigma_{E'}$. So, the commutativity requires the equality
$\sigma_{S'}(t)\otimes \sigma_{E'} = \hat\rho_{S}(t)\otimes
\hat\rho_{E}, \forall{t}$. However, quantum dynamics is continuous
in time. Likewise in Proof of Lemma 1, quantum correlations
relativity guarantees, that for the most of the time instants the
equality will not be fulfilled. Q.E.D.

\label{lastpage-01}

\end{document}